\documentclass{aastex631}

\newcommand{\kms}{\,km\,s$^{-1}$}

\submitjournal{ApJ}

\shorttitle{Morpho-kinematic modeling of the ejecta of V1280 Sco}
\shortauthors{Naito et al.}
\graphicspath{{./}{}}

\begin{document}

\title{Morpho-kinematic modeling of the expanding ejecta of the extremely slow nova V1280 Scorpii}

\correspondingauthor{Hiroyuki Naito}
\email{naito@nayoro-obs.jp}

\author[0000-0001-9067-7653]{Hiroyuki Naito}
\affiliation{Nayoro Observatory, 157-1 Nisshin, Nayoro, Hokkaido 096-0066, Japan}
\affiliation{Graduate School of Science, Nagoya University, Furo-cho, Chikusa-ku, Nagoya, Aichi 464-8602, Japan}

\author[0000-0001-8813-9338]{Akito Tajitsu}
\affiliation{Okayama Astrophysical Observatory, National Astronomical Observatory of Japan, NINS, 3037-5 Honjo, Kamogata, Asakuchi, Okayama 719-0232, Japan}

\author[0000-0003-3617-4400]{Val\'erio A. R. M. Ribeiro}
\affiliation{Instituto de Telecomunica\c{c}\~{o}es, Campus Universit\'{a}rio de Santiago, 3810-193 Aveiro, Portugal}
\affiliation{Departamento de F\'{i}sica, Universidade de Aveiro, Campus Universit\'{a}rio de Santiago, 3810-193 Aveiro, Portugal}
\affiliation{Department of Astrophysics/IMAPP, Radboud University, P.O. Box 9010, 6500 GL Nijmegen, The Netherlands}
\affiliation{Department of Astronomy, University of Cape Town, Private Bag X3, Rondebosch 7701, South Africa}

\author[0000-0002-5756-067X]{Akira Arai}
\affiliation{Subaru Telescope, National Astronomical Observatory of Japan, 650 North A'ohoku Place, Hilo, Hawaii 96720, USA}
\affiliation{Koyama Astronomical Observatory, Kyoto Sangyo University, Motoyama, Kamigamo, Kita-ku, Kyoto, Kyoto 603-8555, Japan}

\author[0000-0003-0332-0811]{Hiroyuki Maehara}
\affiliation{Okayama Astrophysical Observatory, National Astronomical Observatory of Japan, NINS, 3037-5 Honjo, Kamogata, Asakuchi, Okayama 719-0232, Japan}

\author[0000-0002-0008-9979]{Shinjirou Kouzuma}
\affiliation{Faculty of Liberal Arts and Sciences, Chukyo University, 101-2 Yagoto-honmachi, Showa-ku, Nagoya, Aichi 466-8666, Japan}

\author{Takashi Iijima}
\affiliation{Astronomical Observatory of Padova, Asiago Section, Osservatorio Astrofisico, 36012, Asiago (Vi), Italy}

\author[0000-0001-9547-8290]{Atsuo T. Okazaki}
\affiliation{Faculty of Engineering, Hokkai-Gakuen University, Toyohira-ku, Sapporo, Hokkaido 062-8605, Japan}

\author[0000-0002-3656-4081]{Makoto Watanabe}
\affiliation{Department of Applied Physics, Okayama University of Science, Kita-ku, Okayama, Okayama 700-0005, Japan}

\author[0000-0002-7084-0860]{Seiko Takagi}
\affiliation{Department of Cosmosciences, Hokkaido University, Kita 10, Nishi 8, Kita-ku, Sapporo, Hokkaido 060-0810, Japan}

\author{Fumitake Watanabe}
\affiliation{Nayoro Observatory, 157-1 Nisshin, Nayoro, Hokkaido 096-0066, Japan}

\author[0000-0001-7641-5497]{Itsuki  Sakon}
\affiliation{Department of Astronomy, Graduate Schools of Science, University of Tokyo, 7-3-1 Hongo, Bunkyo-ku, Tokyo 113-0033, Japan}

\author{Kozo Sadakane}
\affiliation{Astronomical Institute, Osaka Kyoiku University, Asahigaoka, Kashiwara, Osaka 582-8582, Japan}



\begin{abstract}

Morphology of nova ejecta is essential for fully understanding the physical processes involved in nova eruptions.
We studied the 3D morphology of the expanding ejecta of the extremely slow nova V1280 Sco with a unique light curve.
Synthetic line profile spectra were compared to the observed [O\,{\footnotesize III}] 4959, 5007 and [N\,{\footnotesize II}] 5755 emission line profiles in order to find the best-fit morphology, inclination angle, and maximum expansion velocity of the ejected shell.
We derive the best fitting expansion velocity, inclination, and squeeze as $V_{\rm exp} = 2100^{+100}_{-100}$ \kms, $i = 80^{+1}_{-3}$ deg, and $squ = 1.0^{+0.0}_{-0.1}$ using [O\,{\footnotesize III}] line profiles, and $V_{\rm exp} = 1600^{+100}_{-100}$ \kms, $i = 81^{+2}_{-4}$ deg, and $squ = 1.0^{+0.0}_{-0.1}$ using  [N\,{\footnotesize II}] 5755 line profile.
A high inclination angle is consistent with the observational results showing multiple absorption lines originating from clumpy gases which are produced in dense and slow equatorially focused outflows.
Based on additional observational features such as optical flares near the maximum light and dust formation on V1280 Sco, a model of internal shock interaction between slow ejecta and fast wind proposed for the $\gamma$-ray emission detected in other novae seems to be applicable to this extremely slow and peculiar nova.
Increasing the sample size of novae whose morphology is studied will be helpful in addressing long-standing mysteries in novae such as the dominant energy source to power the optical light at the maximum, optical flares near the maximum, clumpiness of the ejecta, and dust formation.
\end{abstract}

\keywords{novae, cataclysmic variables -- stars: winds, outflows -- stars: mass-loss -- stars: individual: V1280 Sco -- techniques: spectroscopic}


\section{Introduction}
\label{sec:introduction}

A classical nova is caused by a thermonuclear runaway on the surface of a white dwarf (WD) following the accretion of hydrogen-rich material from a non-degenerate stellar companion in a close binary system \citep[e.g.,][]{2003cvs..book.....W, 2008clno.book.....B, 2020AARv..28....3D, 2020arXiv201108751C}. 
Thermal and kinetic energies of the accreted gas material are powered by radiation from the nuclear burning and the envelope greatly expands, increasing its brightness rapidly.
When the photospheric radius of the nova ejecta is greater than the binary separation, a common envelope is formed in which the ejected matter interacts with the secondary star.
After optical peak, the optical luminosity of a nova declines exponentially due to continuous nuclear fusion on the WD surface at a quasi-steady rate.
As most of the accreted material is released by the optically thick wind, the nuclear burning terminates, and then eventually the WD returns to its quiet state.
Such evolution of nova outbursts has been accepted as a standard model  \citep[e.g.,][]{1986ApJ...310..222P, 1994ApJ...437..802K}.

A maximum magnitude versus rate of decline (MMRD) relation, which is an empirical relationship that a faster declining nova shows a brighter optical peak, is still in debate \citep[e.g.,][]{2017ApJ...839..109S, 2018MNRAS.476.4162O, 2018MNRAS.481.3033S, 2020ApJ...902...91H}.
On the theoretical side, however, \citet{2006ApJS..167...59H} discovered a universal decline law of classical novae based on the free-free emission and the optically thick wind theory which reproduced well the optical light curves of many observed novae, especially showing a smooth decline in brightness \citep[e.g.,][]{2010ApJ...709..680H, 2016ApJ...816...26H, 2018ApJS..237....4H, 2019ApJS..241....4H, 2019ApJS..242...18H}.
Their model successfully predicts essential parameters of novae such as the WD mass and the distance under the assumption of spherical symmetry.
According to \citet{2020ApJ...902...91H}, the maximum visual magnitude for specific novae, for example the S-types defined by \citet{2010AJ....140...34S}, can be derived if the ignition mass associated with the mass accretion rate is given.
On the other hand, the optical luminosity of the nova emitting $\gamma$-ray radiation near the optical maximum is proposed to be contributed considerably by shock interactions between distinct fast and slow outflows \citep{2014Natur.514..339C, 2017NatAs...1..697L, 2020NatAs...4..776A, 2020ApJ...905...62A}.
It is suggested that the slow outflow is associated with the initial ejection concentrated in the equatorial plane and the fast flow is likely a radiation pressure driven wind from the WD.
The observations strongly suggest that their samples, and many novae, are multiphased and aspherical \citep[see also, e.g., ][]{2000MNRAS.314..175G, 2014ApJ...785...78N}.
Thus, the mass ejection near maximum light, which is directly connected to the MMRD relation, is still poorly understood owing to the diversity of nova properties. Another parameter that is often not taken into account is the morphology of the ejecta. This should be a key to revealing the physical process of mass loss during the nova eruption and the diversity of novae.

The nova V1280 Scorpii (hereafter, V1280 Sco) was independently discovered by two Japanese amateur astronomers (Y. Nakamura and Y. Sakurai) on 2007 February 4.85 \textsc{\lowercase{UT}} (defined as $\Delta t$ = 0 days, MJD = 54,135.845) at the ninth visual magnitude \citep{2007IAUC.8803....1Y}, which was followed by a somewhat slow rise to its maximum brightness of $V$ = 3.78 mag on February 16.19 \citep{2007CBET..852....1M}.
A low-dispersion optical spectrum obtained on February 5.87 (one day after the discovery) showed Balmer and  Fe\,{\footnotesize II} lines with P Cygni absorption profiles and the object was confirmed as a classical nova \citep{2007IAUC.8803....2N}.
\citet{2012AA...543A..86N} reported the results of photometric and spectroscopic observations from pre-maximum to plateau phase (from 2007 February to 2011 April), summarizing that V1280 Sco is an extremely slow nova which has an exceptionally long plateau spanning over 1000 days. V1280 Sco only entered the nebular phase about 50 months after eruption, which allowed \citet{2012AA...543A..86N} to estimate the mass of the WD to be lower than $\sim$0.6 $M_\mathrm{\sun}$.

\begin{figure}
\epsscale{1.2}
\plotone{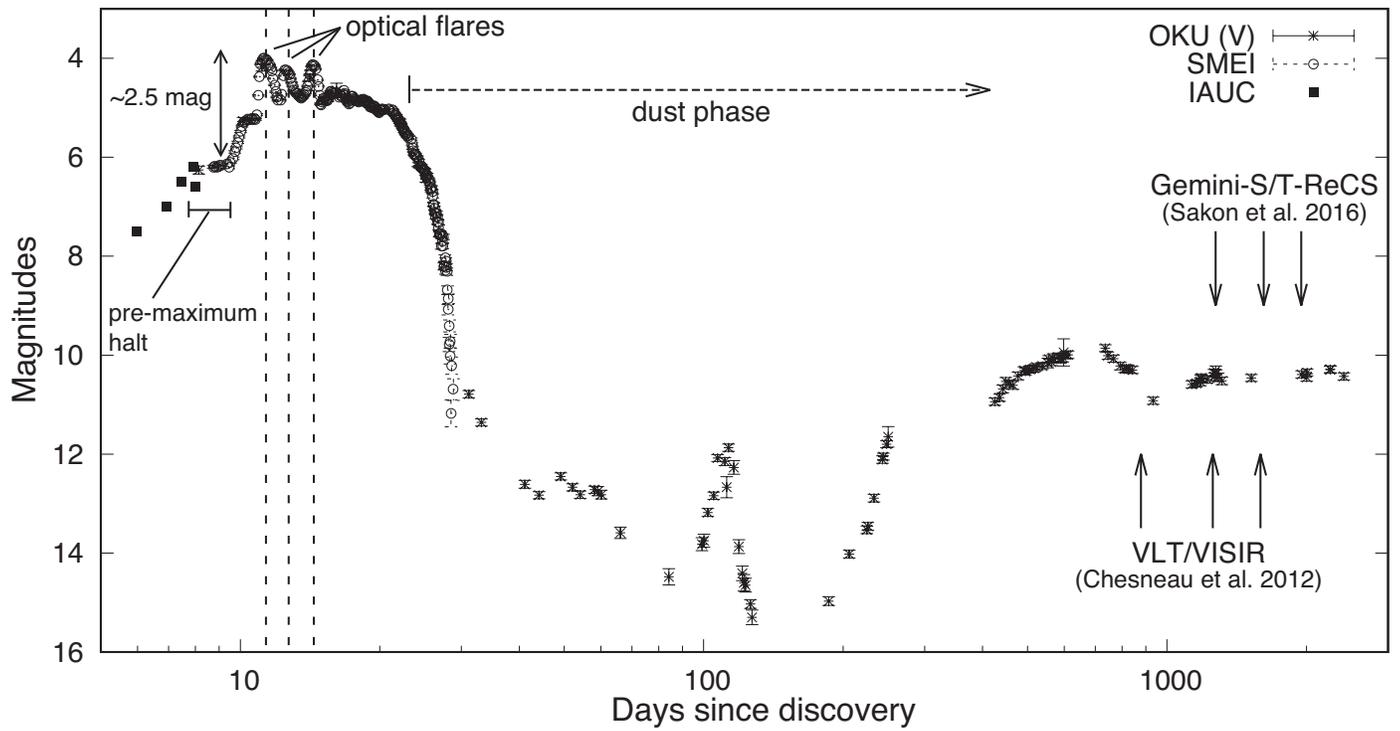}\\
\caption{The SMEI data \citep{2010ApJ...724..480H} are plotted together with OKU $V$-magnitude and the data collected from IAU Circulars \citep{2007IAUC.8803....1Y,2007IAUC.8807....1Y}, showing the pre-maximum halt and the optical flares around the peak brightness.
The arrows indicate the dates of mid-infrared direct imaging observations using the Very Large Telescope with VISIR \citep{2012AA...545A..63C} and using the Gemini South telescope with T-ReCS \citep{2016ApJ...817..145S}.}
\label{fig:LC_SMEI_V_final}
\end{figure}

V1280 Sco can be considered to be one of the most important novae as it has multifaceted properties associated with the long-standing nova mysteries such as multiple variations near the maximum, dust formation, and the clumpiness of the ejecta \citep[e.g.,][]{2010ApJ...724..480H, 2012AA...545A..63C, 2010PASJ...62L...5S}.
Using high time resolution data obtained by the Solar Mass Ejection Imager (SMEI) on board the {\it Coriolis} satellite, \citet{2010ApJ...724..480H} revealed that this nova experiences some short episodes (each time scale is less than one day) of re-brightening with amplitude of $\sim$1 mag near the peak (also see Figure \ref{fig:LC_SMEI_V_final}).
Dust formation around V1280 Sco was reported to occur 11 days after the maximum \citep{2008AA...487..223C} and has been investigated by various instruments in mid- and near-infrared wavelengths \citep[e.g.,][]{2008MNRAS.391.1874D, 2008AA...487..223C, 2012AA...545A..63C, 2016ApJ...817..145S}. The presence of a dusty hourglass-shaped bipolar (or elongated shape) nebula around V1280 Sco was revealed based on high-spatial resolution observations using the Very Large Telescope \citep{2012AA...545A..63C} and the Gemini South telescope \citep{2016ApJ...817..145S} between 2009 and 2012.
Multiple absorption lines likely originating from clumpy gas clouds (ejected shells) were observed in high-dispersion spectra obtained with the 8.2 m Subaru telescope \citep[][]{2010PASJ...62L...5S, 2013PASJ...65...37N}.
Such multiple components of Na\,{\footnotesize I} D, Ca\,{\footnotesize II} H and K, and metastable He\,{\footnotesize I}* had been detected at least until 2012 (five years after the eruption), which stands in contrast to observable lifetimes (2-8 weeks) of the transient heavy element absorption (THEA) systems (e.g., Sc\,{\footnotesize II}, Ti\,{\footnotesize II}, Cr\,{\footnotesize II}, Fe\,{\footnotesize II}) referred to in \citet{2008ApJ...685..451W} and the radioactive isotope of beryllium $^7$Be\,{\footnotesize II} \citep{2015Natur.518..381T, 2016ApJ...818..191T}.
\citet{2020arXiv200711025R}, furthermore, highlighted its uniqueness that V1280 Sco is the only nova that has been at an unusually high and stable level of brightness for more than 10 years after the recovery from the dust extinction.

To seek clues to the cause of these nova phenomena, we investigate the morphology of V1280 Sco using [O\,{\footnotesize III}] and [N\,{\footnotesize II}] forbidden lines in this paper. In Section \ref{sec:observations} our observational data are presented, while in Section \ref{sec:modeling} we discuss our modeling techniques and associated assumption. Section \ref{sec:modeling_results} shows our model fits to the observed line profiles. In Section \ref{sec:discussion}, we discuss our results and finally present the conclusions in Section \ref{sec:conclusions}.

\section{Observations and results}
\label{sec:observations}

\begin{figure*}[ht]
\begin{center}
\epsscale{1.25}
\centering 
\includegraphics[scale=1.5]{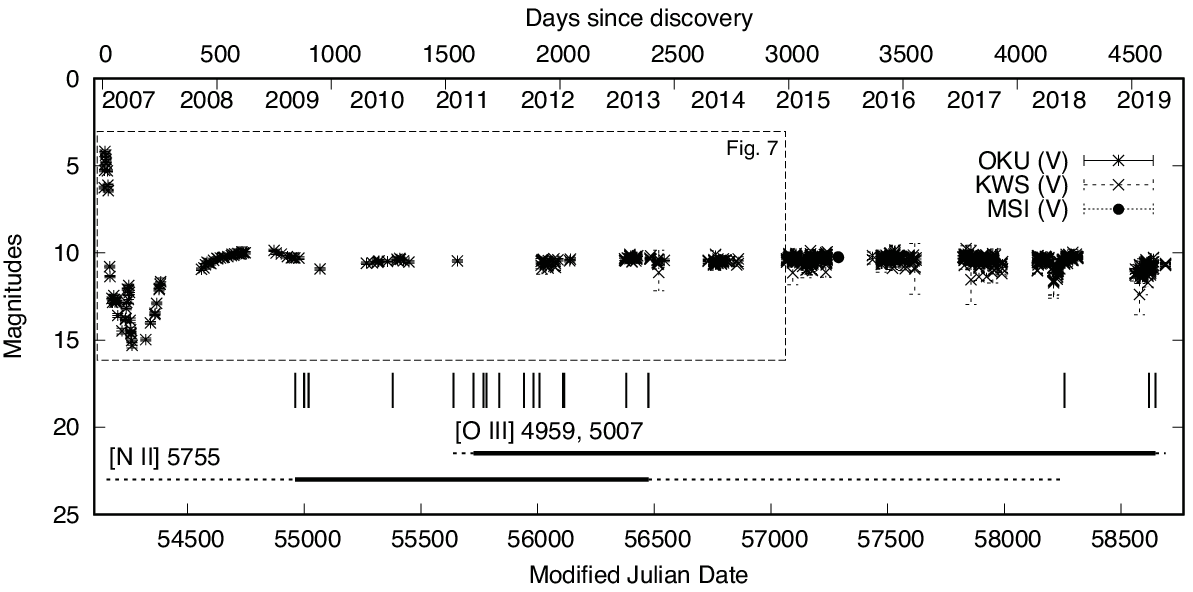}
\caption{The observed $V$-mag light curve, spanning over a decade, for V1280 Sco. The data were collected from Osaka Kyoiku University (OKU), Kamogata/Kiso/Kyoto Wide-field Survey (KWS) and Nayoro Observatory of Hokkaido University (NOHU). Epochs of spectroscopic observations using the 8.2 m Subaru telescope are indicated by vertical lines. Appearance terms for [OIII] 4959, 5007 and [NII] 5755 are corresponding to the thick horizontal lines, while the dashed lines show the period when each appearance is not confirmed due to missing spectral data. There is an exceptionally long plateau spanning over 1000 days. The dashed box exhibits the range of the light curves plotted in Figure \ref{fig:LC_SMEI_V_final} which shows the pre-maximum halt and the optical flares around the peak brightness.}
\label{fig:light_curve_of_V1280Sco}
\end{center}
\end{figure*}

High-dispersion spectroscopic observations of V1280 Sco were conducted with the High Dispersion Spectrograph \citep[HDS;][]{2002PASJ...54..855N} attached to the 8.2 m Subaru telescope from 2009 to 2019.
This observation period corresponds to the long plateau phase spanning over 10 years shown in Figure \ref{fig:light_curve_of_V1280Sco}, where the $V$-mag data from 2012 to 2019 are obtained by Kamogata/Kiso/Kyoto Wide-field Survey (KWS)\footnote{Available from http://kws.cetus-net.org/~maehara/VSdata.py}  (a total of 785 data points) and our photometric observations using the 1.6 m Pirka telescope with Multi-Spectral Imager \citep[MSI;][]{2012SPIE.8446E..2OW} at Nayoro Observatory of Hokkaido University (a total of 10 data points), following the data obtained at Osaka Kyoiku University (OKU) from 2007 to 2011 published in \citet{2012AA...543A..86N}.

\begin{deluxetable*}{lrccccccccl}
\tablecaption{Journal of spectroscopic observations of V1280 Sco using the 8.2 m Subaru telescope.}
\label{table:observation_log}
\tablehead{
\colhead{Date} & \colhead{UT}\tablenotemark{a} & \colhead{MJD} & \colhead{$\Delta{t}$}\tablenotemark{b} & \colhead{Exposure} & \colhead{Range}  & \colhead{Slit width} & \colhead{PA}\tablenotemark{c} & \colhead{ [O\,{\footnotesize III}] 4959, 5007}\tablenotemark{d} & \colhead{ [N\,{\footnotesize II}] 5755}\tablenotemark{d} & \colhead{Remarks} \\
\colhead{} & \colhead{h m} & \colhead{} & \colhead{(days)} & \colhead{time (s)} &  \colhead{(\AA)} & \colhead{(")} & \colhead{(deg)} & \colhead{}  & \colhead{} & \colhead{} 
}\
\startdata
2009 & & & & & & & & & \\
May 9 & 12 24 & 54,960.517 &  825 & 900 & 4100-6860  & 0.6 & 88-93 &  &  \checkmark &\\
Jun 15 & 12 36 & 54,997.525 & 862 & 600 & 4100-6860  &  0.6  & 143-146 &  &  \checkmark & Fig.~\ref{fig:evolution_of_spectra} \\
Jun 16 & 6 45 & 54,998.282 & 863 & 900 &  3400-5110 &  0.6  & 54-55  &  & & \\
Jun 16 & 11 48 & 54,998.492 & 863 & 900 & 4110-6870 &  0.6  & 127-133   &  &  \checkmark & \\
Jun 16 & 12 14 & 54,998.510 & 863 & 900 & 5310-8070  & 0.6  & 136-142 &  &  \checkmark & \\
Jul 4 & 10 51 & 55,016.452 & 881 &900 & 4100-6860 & 0.6  & 132-137 &  &  \checkmark & Fig.~\ref{fig:evolution_of_spectra} \\
Jul 6 &  9 19 & 55,018.389  & 883 &1200 & 4100-6860 & 0.6  & 103-110 &  &  \checkmark  & \\
Jul 6 & 10 37 & 55,018.443  & 883 & 600 & 3400-5110 & 0.6  & 130-134 &  & & \\
\hline
2010 & & & & & & & & & \\
Jul 1&  9 43  & 55,378.405 & 1243 & 1200 & 4130-6860 & 0.6  & 104-111  &  &  \checkmark & Fig.~\ref{fig:evolution_of_spectra} \\
\hline
2011  & & & & & & & & & \\
Mar 17 & 15 27 & 55,637.644 & 1502 & 930 & 4100-6860 &  0.6  & 80-85 &  &  \checkmark & Fig.~\ref{fig:evolution_of_spectra} \\
Jun 12 & 10 33 & 55,724.440  & 1589 & 1200 & 4100-6860 & 0.6  & 95-102 &  &  \checkmark & Figs.~\ref{fig:evolution_of_spectra},  \ref{fig:modeling_results_[NII]} \\
Jul 24 & 6 36 &  55,766.275 & 1631 & 1800 & 4100-6860 & 0.6  & 74-82 &  \checkmark &  \checkmark & Fig.~\ref{fig:evolution_of_spectra}\\
Aug 6  & 5 43 & 55,779.239 & 1644 & 1200 & 4100-6860  & 0.6  & 73-79 &  \checkmark &  \checkmark  & \\
Aug 6  & 6 39 & 55,779.277 & 1644 & 1200 & 3030-4630 &  0.6  & 89-96 &  & &  \\
Sep 30  & 5 39 & 55,834.236 & 1699 & 2400 & 4130-6860 &  0.6  & 143-157 &  \checkmark &  \checkmark & Fig.~\ref{fig:evolution_of_spectra} \\
\hline
2012  & & & & & & & & & \\
Jan 13 & 16 17 & 55,939.679 & 1804 & 930  & 5080-7850 &  4.0  & 53-53 &  &  \checkmark & \\
Feb 24 & 15 29 & 55,981.645 & 1846 & 1800 &  4100-6860 &  0.6  & 62-68 &  \checkmark &  \checkmark & Fig.~\ref{fig:modeling_results_[OIII]}\\
Mar 20  & 14 41 & 56,006.612 & 1871 & 1200 &  4100-6860 &  0.6  & 72-77 &   \checkmark &  \checkmark & Fig.~\ref{fig:evolution_of_spectra}\\
Mar 20  & 15 14 & 56,006.635 & 1871 & 1800 & 3400-5110 & 0.6  & 81-90 &  \checkmark & & \\
Jun 30  & 7 53 & 56,108.328 & 1973 & 1200 & 3540-5250 & 4.0 & 70-75 &  \checkmark & & \\
Jul 4  & 10 \,\,\,3 & 56,112.419 & 1977 & 1200 & 4100-6860 & 4.0 & 116-123 &   \checkmark &  \checkmark & \\
\hline
2013  & & & & & & & & & \\
Mar 27 & 13 42 & 56,378.571 & 2243 & 900 & 4110-6860 &  0.6  & 65-68 &   \checkmark &  \checkmark &Fig.~\ref{fig:evolution_of_spectra} \\
Jun 30 & 9 34 & 56,473.399 & 2338 & 900 & 6130-8890 & 4.0 & 100-105 &  & & \\
Jul 1  & 9 \,\,\,5 & 56,474.379 & 2339 & 900 & 4110-6770 & 4.0 & 91-96 &   \checkmark &  \checkmark & \\
Jul 1  & 9 53 & 56,474.412 & 2339  & 900 & 3040-4630& 0.6  & 108-113  &  & &\\
\hline
2018  & & & & & & & & & \\
May 18 & 11 \,\,\,4 & 58,256.461 & 4121 & 600 & 4110-6860 &  0.6  & 75-78 &  \checkmark & & Fig.~\ref{fig:evolution_of_spectra} \\
\hline
2019  & & & & & & & & & \\
May 15  & 14 20 & 58,618.598 & 4483 & 1200 & 3520-6230 &  0.6  & 136-143 &  & & \\
May 15  & 14 48 & 58,618.617 & 4483 & 1200 & 4110-6860 &  0.6  & 146-156 &  & & Fig.~\ref{fig:evolution_of_spectra} \\
Jun 12 & 11 25 & 58,646.476 & 4511 & 1200 & 4030-6780 & 0.6  & 113-120 &  & & \\
\enddata
\tablenotetext{a}{Universal time at the start time of an exposure.}
\tablenotetext{b}{Days after the discovery (from 2007 February 4). }
\tablenotetext{c}{Position angle (PA) range during the exposure.}
\tablenotetext{d}{Checkmarks are spectra used for the analysis in Section \ref{sec:modeling}.}
\end{deluxetable*}

\begin{figure}
\epsscale{1.2}
\plotone{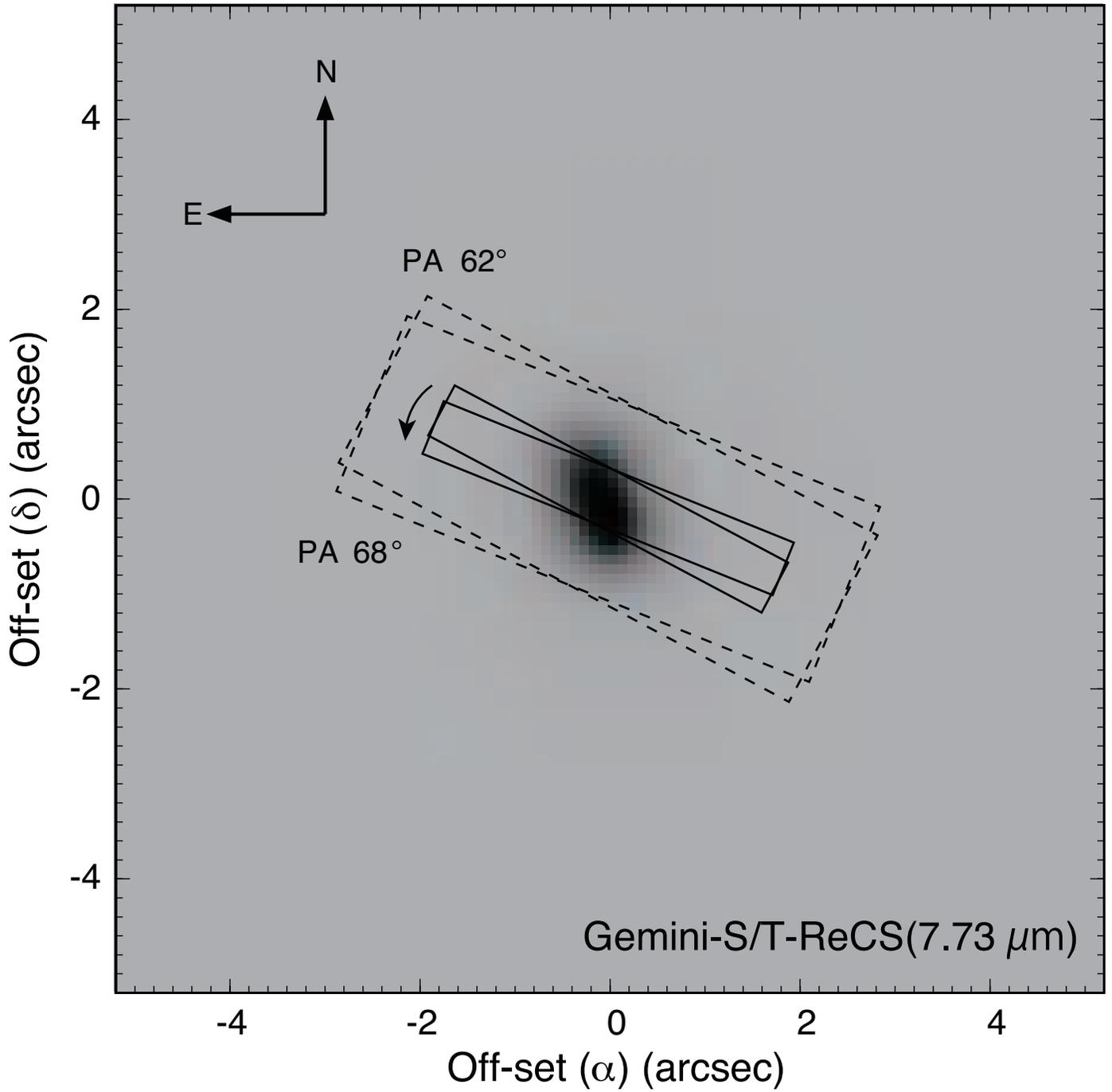}\\
\caption{Subaru/HDS slit position on 2012 Feb. 24 ($\Delta{t}$ = 1846 d), for an example. North is up and east is to the left. The HDS slit position is overlaid on the Gemini South/T-ReCS mid-infrared image (7.73 $\mu$m), showing a bright inner dusty component, taken on 2012 Jun. 6 ($\Delta{t}$ = 1949 d). Dashed line shows the expected actually observed region due to seeing effect ($\sim1"$) and error of the guiding without image rotator ($\sim1"$) .}
\label{fig:V1280Sco_Gemini_final}
\end{figure}

Table~\ref{table:observation_log} provides a journal of spectroscopic observations.
We obtained 29 spectra under nine configurations (different wavelength ranges) of the spectrograph, which covered the wavelength region from 3030 \AA\,  to 8890 \AA\, in total.
A typical resolving power is $R$ $\sim$60,000 at H$\alpha$, while the slit width is 0.3 mm (corresponding to 0".6 in the sky). Each position angle (PA) of the slit was not controlled with an image rotater during the exposure. Figure \ref{fig:V1280Sco_Gemini_final} shows, for example, the Subaru/HDS slit position on 2012 Feb. 24 ($\Delta{t}$ = 1846 d), overlaid on the mid-infrared image taken on 2012 Jun. 6 ($\Delta{t}$ = 1949 d) using the Gemini South telescope with T-ReCS (also see Sect. 5.1).

Data reduction was carried out using the \textsc{\lowercase{IRAF}}\footnote{\textsc{\lowercase{IRAF}} is distributed by the National Optical Astronomy Observatory, which is operated by the Association of Universities for Research in Astronomy (AURA) under a cooperative agreement with the National Science Foundation \citep[e.g.,][]{1986SPIE..627..733T}.} software package in the standard manner. Wavelength calibration was performed using the Th-Ar comparison spectrum obtained during each observation.
All spectra were converted to the heliocentric scale.
Reddening correction is not performed.

\begin{figure}
\epsscale{0.75}
\plotone{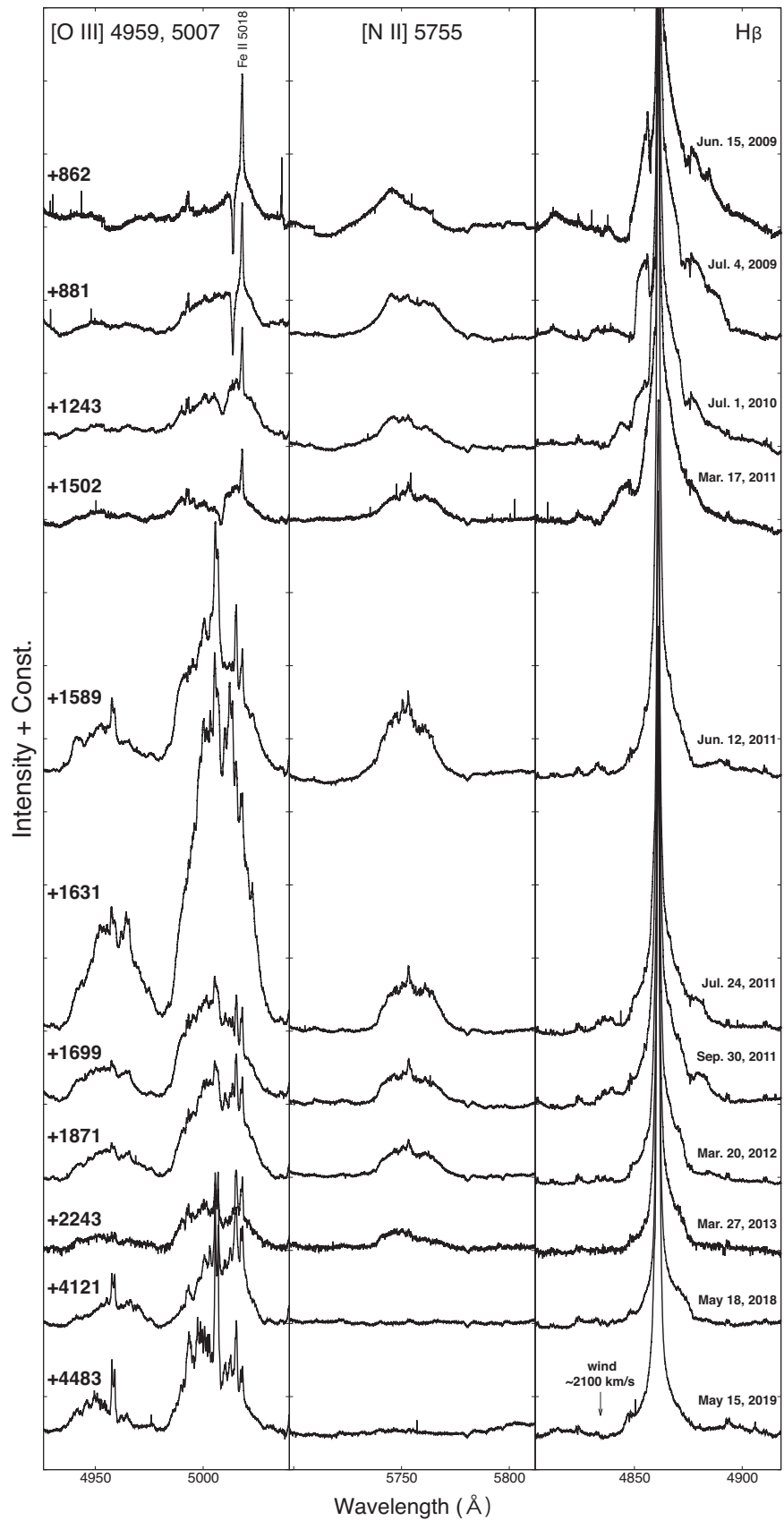}\\
\caption{Evolution of [O\,{\footnotesize III}] 4959, 5007 and  [N\,{\footnotesize II}] 5755 forbidden lines and H$\beta$ recombination line. Number on the left are days after the discovery.}
\label{fig:evolution_of_spectra}
\end{figure}

Figure \ref{fig:evolution_of_spectra} shows the evolution of [O\,{\footnotesize III}] 4959, 5007 and  [N\,{\footnotesize II}] 5755 forbidden lines and H$\beta$ recombination line using the selected spectra (see remarks in Table~\ref{table:observation_log}).
Their appearances of [O\,{\footnotesize III}] 4959, 5007 and [N\,{\footnotesize II}] 5755 are shown as thick horizontal lines in Figure \ref{fig:light_curve_of_V1280Sco}, while dashed lines show the period when each appearance is not confirmed due to missing spectral data.
V1280 Sco had taken a very long time ($\sim$50 months) to enter the nebular phase, according to a clear detection of both [O\,{\footnotesize III}] 4959 and 5007.
There appeared to be P Cygni like absorptions on H$\beta$ lines  from 2009 to 2011.
The nova continues generating the wind ($\sim$2100 \kms) caused by the hydrogen burning at least for about 4500 days until 2019.

\section{Modeling}
\label{sec:modeling}

Optical images of resolved nova remnants show that nova shells tend to be clumpy and have a myriad of structures \citep[e.g.,][]{1972MNRAS.158..177H, 1983ApJ...273..647S, 2000MNRAS.314..175G, 2003MNRAS.344.1219H, 2020ApJ...892...60S}.
A correlation appears to exist between the speed class of the nova and the degree of shaping, suggesting the slower the nova, the more prolate the nebular remnant \citep{1995MNRAS.276..353S}. 
An interpretation is given by \citet{1997MNRAS.284..137L} that the effects of the interaction of the ejecta during a common envelope phase vary among novae of different speed class, which are associated to the different time-scales of the slow and fast wind phases. Here, we aim to disentangle the morphology of V1280 Sco as an extremely slow nova by means of morpho-kinematical studies of the [O\,{\footnotesize III}]  and  [N\,{\footnotesize II}] line profiles.

In order to disentangle the 3D geometry and kinematic structure of V1280 Sco we used \textsc{\lowercase{SHAPE}}\footnote{Available from http://www.astrosen.unam.mx/shape/}\citep{2011ITVCG..17..454S}, morpho-kinematic modeling and reconstruction tool for astrophysical objects, which have yielded various results for nova shells \citep[e.g.,][]{2009ApJ...703.1955R, 2011MNRAS.410..525M, 2011MNRAS.412.1701R, 2013MNRAS.433.1991R, 2013ApJ...768...49R, 2018AA...611A...3H, 2020MNRAS.495.2075P, 2022MNRAS.512.2003S}.
\textsc{\lowercase{SHAPE}} is particularly suited to study expanding nova shells showing [O\,{\footnotesize III}]  and  [N\,{\footnotesize II}] forbidden lines like V1280 Sco as it has been developed for modeling optically thin environments (e.g., planetary nebulae).

We modeled our observations assuming a bipolar geometry, where we varied the inclination angles, maximum expansion velocities and squeezes.
A total of 72,800 synthetic spectra made with different parameter combinations of morphologies, inclination angles  (where an inclination $i$ = $90^\circ$ corresponds to the orbital plane being edge-on) from $0^\circ$ to $90^\circ$ (in steps of $1^\circ$), and maximum expansion velocities ($V_{\rm exp}$) from 100 and 8000 \kms (in steps of 100 \kms) assumed to take place in a Hubble flow where material furthest from the binary system is moving fastest.
We then flux matched the synthetic and observed spectra by $\chi^{2}$ minimization analysis.
Additionally we introduced two new parameters: i)  the ratio of  [O\,{\footnotesize III}] 4959 to 5007 to take into account the difference of interstellar reddening effect between [O\,{\footnotesize III}] 4959 and [O\,{\footnotesize III}] 5007, and ii) a shift velocity to correct the Doppler effect due to relative motion between the object and observer.
The assumed bipolar structure is based on the previous study for Nova Mon 2012 (V959 Mon), where the simplest morphology of [O\,{\footnotesize III}] 4959, 5007 is found to be that of a bipolar structure \citep{2013ApJ...768...49R}.
Forbidden lines such as [O\,{\footnotesize III}] and [N\,{\footnotesize II}] come from the outer regions, which tend to be broadly symmetric profiles and trace well the expansion velocity and geometry of the ejecta, wheres Balmer lines come from the whole of the ionized ejecta, a part of which suffers some absorption causing asymmetry represented by P Cygni profile.
As noted in Section \ref{sec:observations}, the line profiles of  [O\,{\footnotesize III}]  and  [N\,{\footnotesize II}] are quite different from that of H$\beta$, which suggests that these origins are also different.

\begin{deluxetable*}{lcccccccc}
\caption{The best fit parameter values for [O\,{\footnotesize III}] 4959, 5007 and [N\,{\footnotesize II}] 5755 analyses.}
\label{tab:best_fit_table}
\tablehead{
\colhead{Date} & \colhead{MJD} & \colhead{$V_{\rm exp}$} & \colhead{Inclination} & \colhead{Squeeze} & \colhead{ [O\,{\footnotesize III}] 5007/4959} & \colhead{ Shift velocity} & \colhead{$\chi^{2}$ /} & \colhead{$P$-value}\\
\colhead{} & \colhead{} & \colhead{(\kms)} & \colhead{(deg)} & \colhead{} & \colhead{ratio} & \colhead{(\kms)}  & \colhead{dof}  & \colhead{}
}\
\startdata
\multicolumn{9}{c}{[O\,{\footnotesize III}] 4959, 5007}\\
\hline
2011-07-24 & 55,766.275 & $1900^{+0}_{-0}$& $78^{+0}_{-0}$ & $1.0^{+0.0}_{-0.0}$ & 3.2 & 65 & 0.77946 & 0.9779 \\
2011-08-06 & 55,779.239 & $2000^{+0}_{-0}$ & $77^{+1}_{-0}$ & $0.9^{+0.0}_{-0.0}$ & 3.7 & 100 & 0.80622 & 0.9593 \\
2011-09-30 & 55,834.236 & $2200^{+0}_{-0}$ & $81^{+0}_{-1}$ & $1.0^{+0.0}_{-0.0}$ & 3.5 & 50 & 1.0003 & 0.4836 \\
2012-02-24 & 55,981.645  & $2100^{+100}_{-100}$ & $80^{+1}_{-3}$ & $1.0^{+0.0}_{-0.1} $ & 3.8 & 90 & 0.62846 & 0.9999 \\
2012-03-20 & 56,006.612 & $2200^{+0}_{-100}$ & $81^{+0}_{-1}$ & $1.0^{+0.0}_{-0.0}$ & 3.8 & 50 & 0.72668 & 0.9948 \\
2012-03-20 & 56,006.635 & $2100^{+0}_{-100}$ & $81^{+2}_{-1}$ & $1.0^{+0.0}_{-0.0}$ & 3.7 & 50 &  0.72446 & 0.9951 \\
2012-06-30 & 56,108.328 & $2200^{+0}_{-100}$ & $83^{+1}_{-2}$ & $1.0^{+0.0}_{-0.0}$ & 3.8 & 50 & 0.95245 & 0.6451 \\
2012-07-04 & 56,112.419 & $2200^{+0}_{-0}$ & $81^{+0}_{-0}$ & $1.0^{+0.0}_{-0.0}$ & 3.8 & 50 & 0.71391 & 0.9965 \\
2013-03-27 & 56,378.571 & $2100^{+300}_{-0}$ & $81^{+3}_{-3}$ & $0.8^{+0.2}_{-0.0}$  & 3.8 & 50 & 0.69979 & 0.9977 \\
2013-07-01 & 56,474.379 & $2000^{+100}_{-0}$ & $81^{+0}_{-2}$ & $0.8^{+0.1}_{-0.0}$ & 3.8 & 85 & 0.97178 & 0.5809 \\
2018-05-18 & 58,256.461 & $1900^{+100}_{-100}$ & $81^{+1}_{-2}$ & $1.0^{+0.0}_{-0.0}$ & 3.6 & 100 & 0.71925 &0.9954 \\
\hline
\multicolumn{9}{c}{[N\,{\footnotesize II}] 5755}\\
\hline
2009-05-09 & 54,960.517 & $1600^{+300}_{-700}$ & $82^{+5}_{-22}$ & $1.0^{+0.0}_{-0.6}$ & & 50 & 1.0157 & 0.4426  \\
2009-06-15 & 54,997.525 & $1500^{+400}_{-500}$ & $80^{+6}_{-23}$ & $0.8^{+0.2}_{-0.4}$ & & 50 & 1.1388 & 0.2308 \\
2009-06-16 & 54,998.492 & $2000^{+200}_{-600}$ & $79^{+5}_{-15}$ & $0.9^{+0.1}_{-0.4}$ & & 50 & 1.0031&  0.4696 \\
2009-06-16 & 54,998.510 & $1900^{+200}_{-200}$ & $79^{+2}_{-4}$ & $0.9^{+0.1}_{-0.1}$ & & 50 & 0.77003 & 0.9178 \\
2009-07-04 & 55,016.452 & $1800^{+100}_{-0}$ & $80^{+2}_{-1}$ & $0.9^{+0.1}_{-0.0}$ & & 50 & 0.7724 &  0.9118 \\
2009-07-06 & 55,018.389 & $1800^{+100}_{-200}$ & $81^{+1}_{-3}$ & $1.0^{+0.0}_{-0.2}$ & & 50 & 0.73702 & 0.9417 \\
2010-07-01 & 55,378.405 & $1700^{+200}_{-100}$ & $81^{+2}_{-4}$ & $1.0^{+0.0}_{-0.1}$ & & 50 & 0.74756 & 0.9293 \\
2011-03-17 & 55,637.644 & $1800^{+0}_{-100}$ & $79^{+2}_{-1}$ & $1.0^{+0.0}_{-0.0}$ & & 60 & 0.89937 & 0.7037 \\
2011-06-12 & 55,724.440 & $1600^{+100}_{-100}$ & $81^{+2}_{-4}$ & $1.0^{+0.0}_{-0.1}$ & & 50 & 0.47080 & 0.9999 \\
2011-07-24 & 55,766.275 & $1500^{+100}_{-0}$ & $83^{+1}_{-2}$ & $1.0^{+0.0}_{-0.0}$ & & 60 & 0.54222 & 0.9981 \\
2011-08-06 & 55,779.239 & $1600^{+0}_{-100}$ & $83^{+1}_{-2}$ & $1.0^{+0.0}_{-0.0}$ & & 50 & 0.87737 & 0.7346 \\
2011-09-30 & 55,834.236 & $1700^{+0}_{-0}$ & $83^{+1}_{-2}$ & $1.0^{+0.0}_{-0.0}$ & & 50 & 0.87234 & 0.7587 \\
2012-01-13 & 55,939.679 & $1600^{+0}_{-100}$ & $83^{+1}_{-4}$ & $1.0^{+0.0}_{-0.1}$ & & 100 & 0.54133 & 0.9986 \\
2012-02-24 & 55,981.645 & $1600^{+100}_{-700}$ & $83^{+1}_{-40}$ & $1.0^{+0.0}_{-0.2}$ & & 65 & 1.1818 &  0.1583 \\
2012-03-20 & 56,006.612 & $1600^{+100}_{-0}$ & $83^{+0}_{-3}$ & $1.0^{+0.0}_{-0.0}$ & & 50 & 0.68317 & 0.9754 \\
2012-07-04 & 56,112.419 & $1600^{+0}_{-100}$ & $81^{+3}_{-4}$ & $1.0^{+0.0}_{-0.1}$ & & 50 & 0.58529 & 0.9970 \\
2013-03-27 & 56,378.571 & $1800^{+100}_{-200}$ & $85^{+2}_{-5}$ & $1.0^{+0.0}_{-0.1}$ & & 50 & 0.54641 & 0.9988 \\
2013-07-01 & 56,474.379 & $1600^{+100}_{-800}$ & $83^{+7}_{-47}$ & $1.0^{+0.0}_{-0.8}$ & & 50 & 0.97773 & 0.5195 \\
\enddata
\tablecomments{The errors are given for reduced $\chi^{2}$ greater than 1$\sigma$.}
\end{deluxetable*}

\begin{figure}
\epsscale{1.2}
\plotone{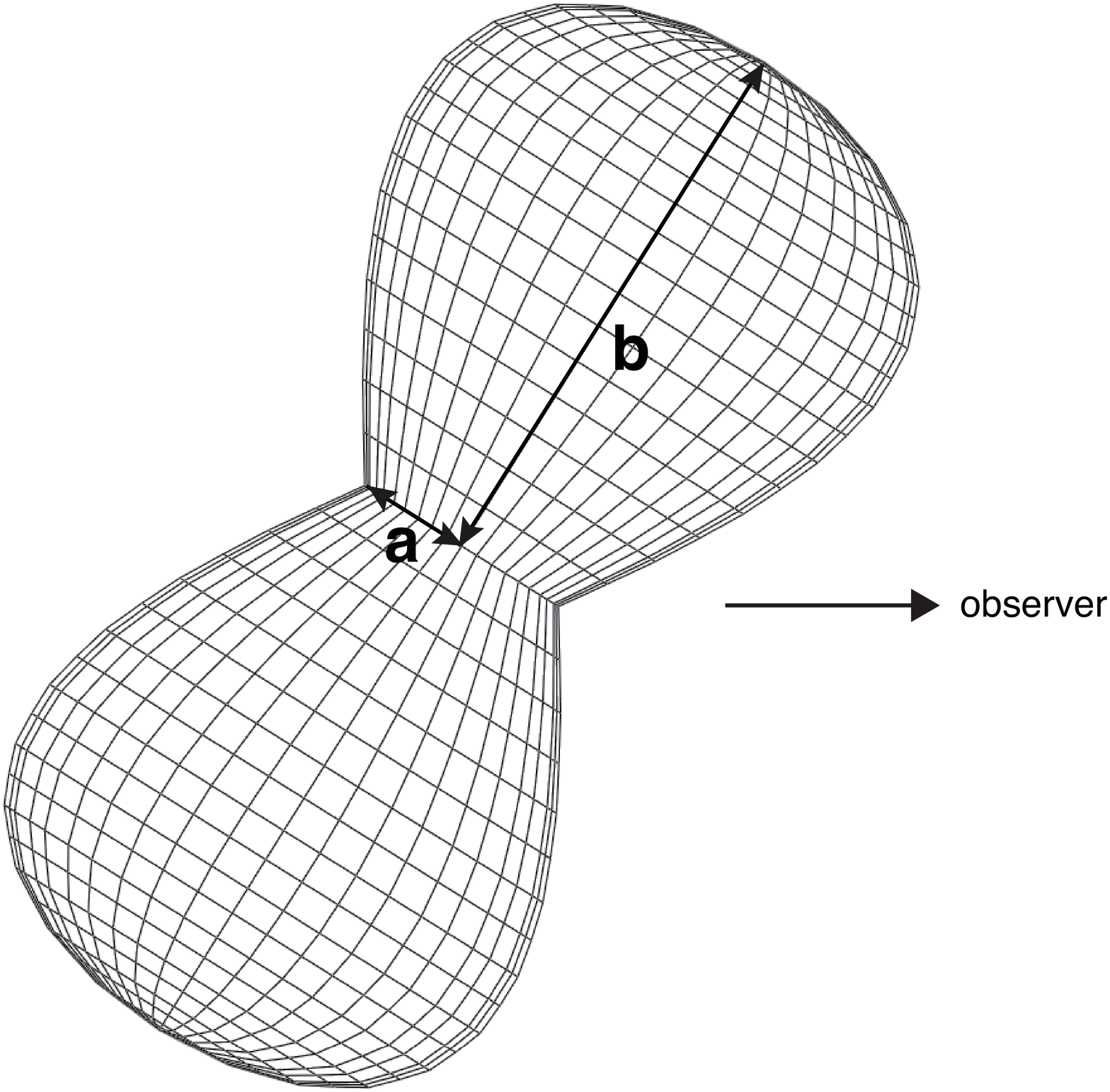}\\
\caption{V1280 Sco model structure ([O\,{\footnotesize III}]  and  [N\,{\footnotesize II}] region) as visualized in \textsc{\lowercase{SHAPE}}. The letters $a$ and $b$ represent the semi-minor and semi-major axes, respectively. The ratio of the semi-minor to semi-major axes defines the squeeze, see Equation (\ref{eq:eq1}). The inclination angle of the binary system is defined as the angle between the plane of the sky and the central binary system's orbital plane. The arrow indicates the observer's direction.}
\label{fig:shape}
\end{figure}

The full parameter space for inclination, maximum expansion velocity, degree of shaping (squeeze from 0.1 to 1.0 in steps of 0.1, defined below), [O\,{\footnotesize III}] 5007/4959 ratio (from 3.0 to 3.8 in steps of 0.1, only for analyses of [O\,{\footnotesize III}] lines), and shift velocity (from $50$ \kms\, to $100$ \kms\, in steps of 5 \kms) were explored to retrieve the synthetic emission line spectrum to compare with the observed spectra to determine $\chi^{2}$.
The squeeze (a modifier within \textsc{\lowercase{SHAPE}}) compresses or expands a structure as a function of position along the symmetry axis (in this case the major axis). We use the fractional amount by which the object is compressed to the squeeze axis (in this case the minor axis).
The squeeze (noted $squ$ as a parameter) is defined as
\begin{equation}
     squ = 1 - \frac{a}{b},
	\label{eq:eq1}
\end{equation}
where $a$ and $b$ are the semi-minor and semi-major axes, respectively, of the ejected shell distance to the center of the binary system (Figure \ref{fig:shape}).

\section{Modeling Results}
\label{sec:modeling_results}

\begin{figure}
\epsscale{1.0}
\plotone{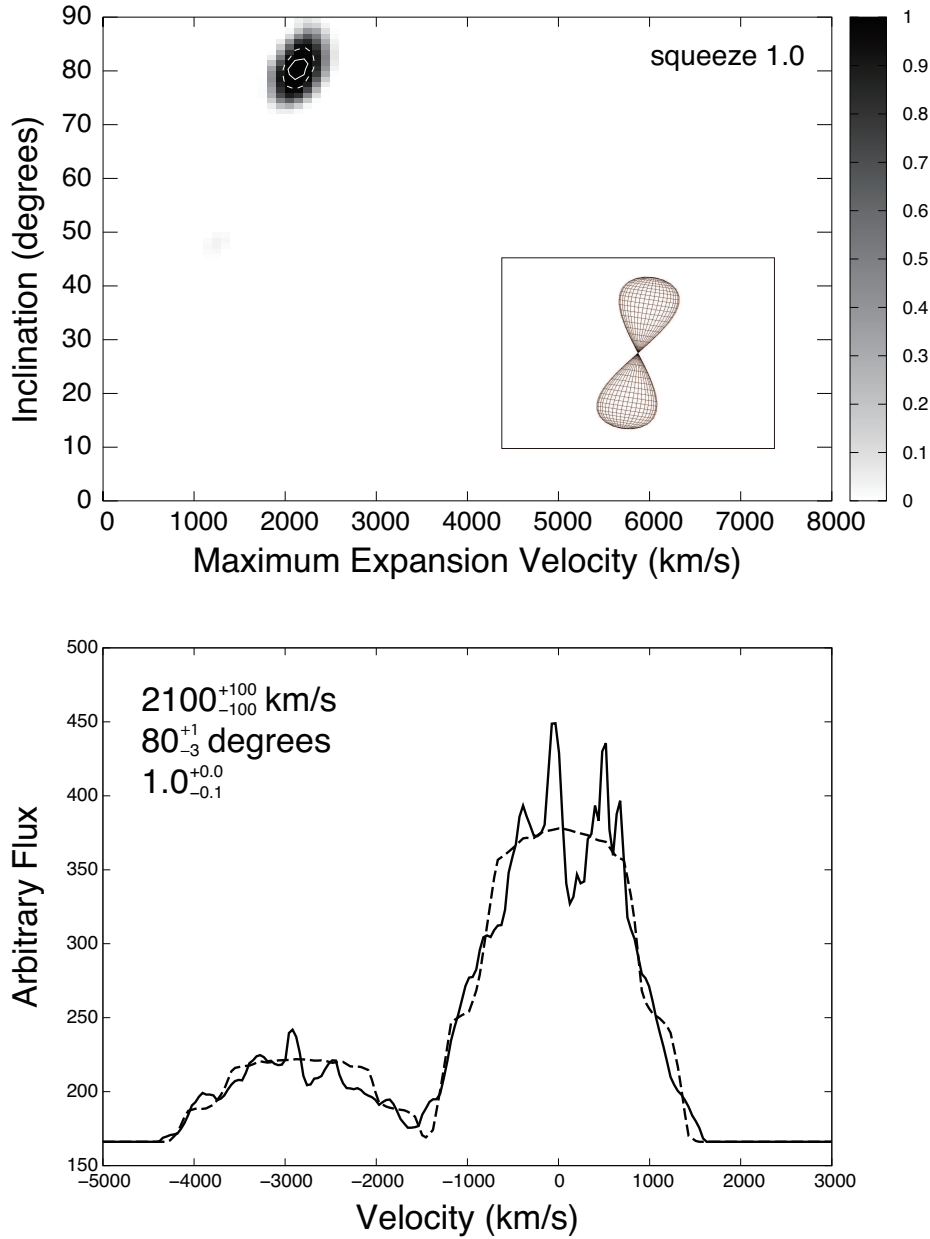}\\
\caption{Top: contour plot showing goodness of fit (reduced $\chi^{2}$) of shape model fits under various assumptions for the inclination and the maximum expansion velocity at squeeze = 1.0 for [O\,{\footnotesize III}] 4959, 5007 lines on February 24, 2012 ($\Delta{t}$ = 1846 d). The solid and dashed (white) lines represent the 1$\sigma$ and 3$\sigma$ boundaries, respectively. The legend bar gives the value of reduced $\chi^{2}$ ranging from 0 to 1. The inset image shows the model structure constructed with the best fit parameters in SHAPE. Bottom: the observed (solid black) and model (dashed black) spectra for the best-fitting parameters (velocity, inclination, and squeeze from top to bottom), shown in the top-left corner with its respective 1$\sigma$ errors. The horizontal axis is heliocentric velocity corresponding to [O\,{\footnotesize III}] 5007. }
\label{fig:modeling_results_[OIII]}
\end{figure}

\begin{figure}
\epsscale{1.0}
\plotone{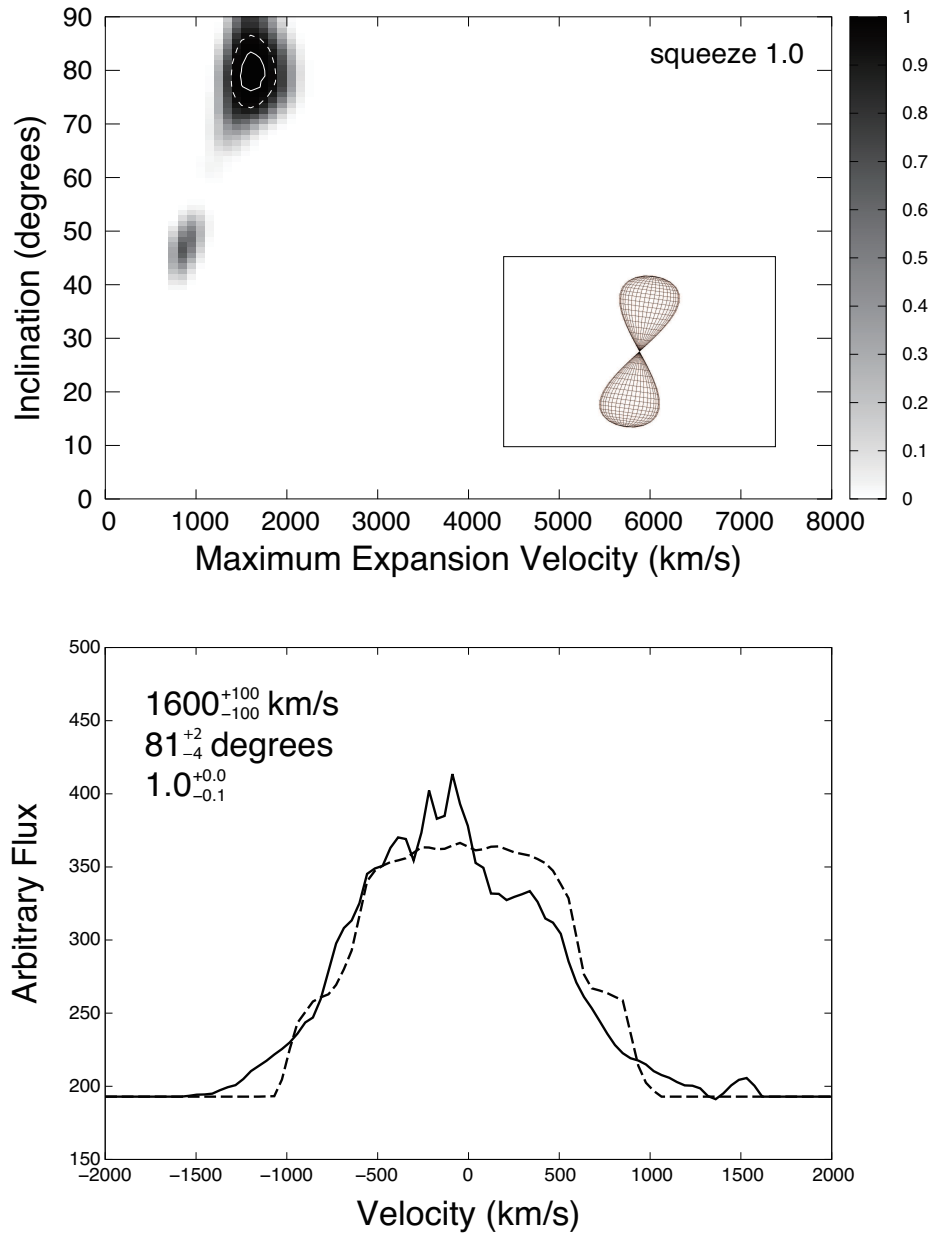}\\
\caption{Same as Figure 4, but for [N\,{\footnotesize II}] 5755 line on June 12, 2011 ($\Delta{t}$ = 1589 d).}
\label{fig:modeling_results_[NII]}
\end{figure}

The results of exploring the full parameter space for [O\,{\footnotesize III}] and [N\,{\footnotesize II}] profiles are shown in Table \ref{tab:best_fit_table}, where the best-fit parameter values for the maximum expansion velocity, inclination angle, and squeeze along with their respective $\chi^{2}$/degree of freedom (dof) (or reduced $\chi^{2}$) and the $P$-value are given. The errors are determined assuming a reduced $\chi^{2}$ greater than 1$\sigma$.
The observed spectra and the model spectra for each of the inclinations, velocities, and squeezes were compared to find the best fit via a $\chi^{2}$ test for  [O\,{\footnotesize III}] and [N\,{\footnotesize II}] emission lines (top of Figures \ref{fig:modeling_results_[OIII]} and \ref{fig:modeling_results_[NII]}, respectively).
We derive the best fitting expansion velocity, inclination, and squeeze as $V_{\rm exp} = 2100^{+100}_{-100}$ \kms, $i = 80^{+1}_{-3}$ deg, and $squ = 1.0^{+0.0}_{-0.1}$ for [O\,{\footnotesize III}] 4959, 5007 on February 24, 2012 (Figure \ref{fig:modeling_results_[OIII]}, bottom), $V_{\rm exp} = 1600^{+100}_{-100}$ \kms, $i = 81^{+2}_{-4}$ deg, and $squ = 1.0^{+0.0}_{-0.1}$ for [N\,{\footnotesize II}] 5755 on June 12, 2011 (Figure \ref{fig:modeling_results_[NII]}, bottom), respectively.

Shown in Table \ref{tab:best_fit_table} implicates that the best-fit model spectra for each are those with high values of inclination and squeeze, replicating well the general features of observed spectra.
It should be kept in mind here that in all cases, such best-fit spectra would not fully replicate the observed spectra because an observed spectrum includes other element lines and/or components originating in different regions, which are not taken into account in the modeling (see bottom of Figure \ref{fig:modeling_results_[OIII]} for example).
As also shown in Figure \ref{fig:evolution_of_spectra}, the contributions from the residuals can change relatively along with the strength of [O\,{\footnotesize III}] and [N\,{\footnotesize II}] lines.
This effect appears as errors in fitting parameter, especially for earlier and later observations of [N\,{\footnotesize II}] 5755 when the line strength is relatively weak.
On the whole, the inclinations and squeezes derived from [O\,{\footnotesize III}] analyses are consistent within the errors with those from using [N\,{\footnotesize II}], while expansion velocities seem to be significantly different between them.
This is likely associated with the difference of ionization region between [O\,{\footnotesize III}] and [N\,{\footnotesize II}], that is, [N\,{\footnotesize II}] 5755 with a higher excitation energy ($\sim$4.05 eV) comes from the inner region than [O\,{\footnotesize III}] 4959, 5007 with a lower excitation energy ($\sim$2.51 eV) when assuming a Hubble flow.
Accordingly, the fact that we derive very similar parameters using the two forbidden lines of [O\,{\footnotesize III}] and [N\,{\footnotesize II}] makes the results more robust.
It is noteworthy that the spectra in each phase are well reproduced by an inclination angle of $\sim$80$^{\circ}$ and a squeeze of $\sim$1.0.

\section{Discussion}
\label{sec:discussion}

\subsection{Infrared images and optical morphology}
\label{sec:Infrared images and optical morphology}
\begin{figure*}[ht]
\epsscale{1.15}
\plotone{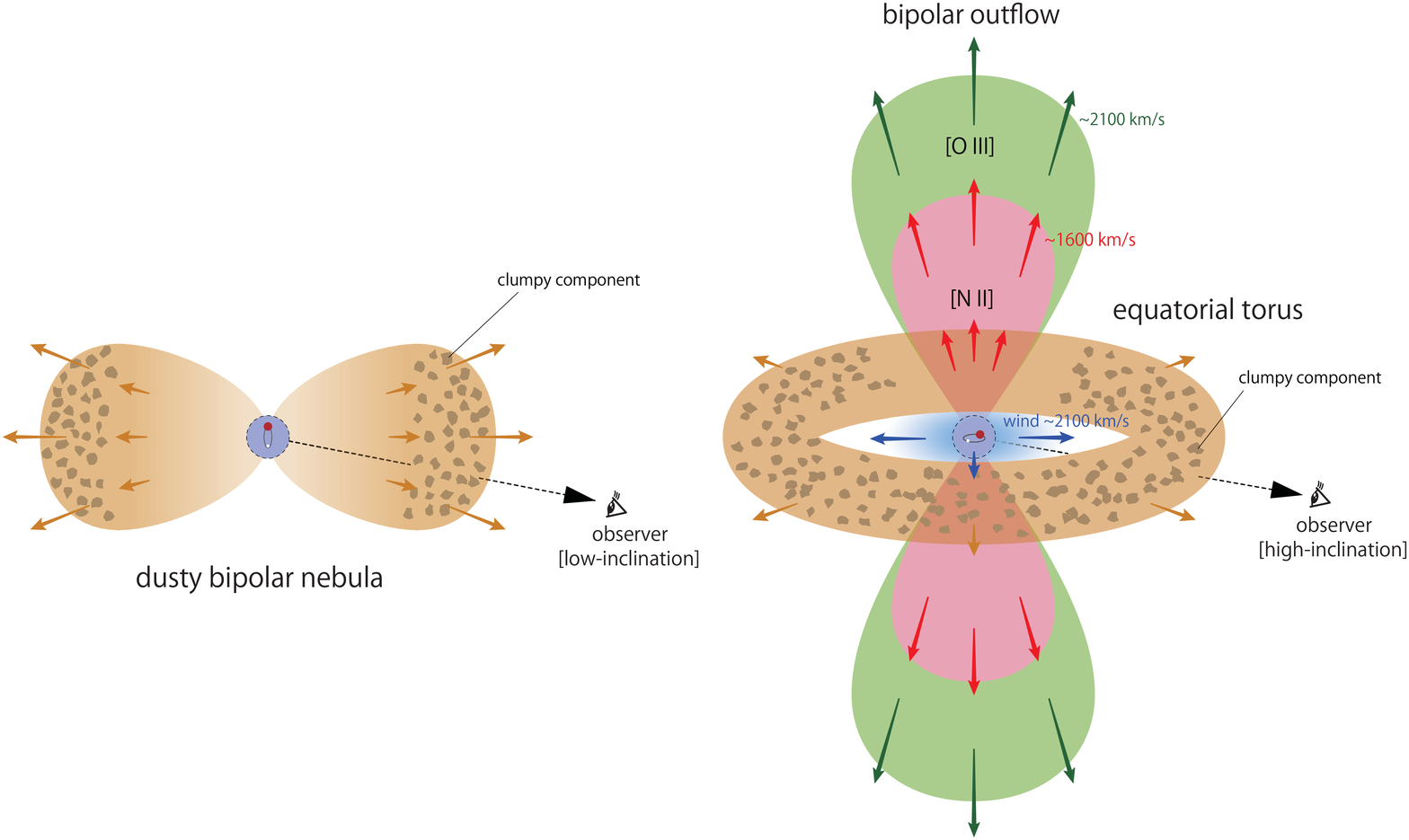}\\
\caption{Left: A schematic of V1280 Sco showing a low-inclination (face-on binary) based on the observations of a dusty hourglass-shaped bipolar (or elongated shape) nebula \citep{2012AA...545A..63C, 2016ApJ...817..145S} and multiple absorption lines \citep{2010PASJ...62L...5S, 2013PASJ...65...37N}, implying a large part of ejected mass is included in polar direction \citep{2013PASJ...65...37N}. Right: A revised schematic of V1280 Sco proposed in this study, illustrating fast wind ([O\,{\footnotesize III}]  and  [N\,{\footnotesize II}] forbidden line regions) extended in polar direction and slow clumpy absorption components ejected in equatorial direction. The origin of the fast flow is likely a radiation-driven wind from the continuous nuclear burning on the surface of the WD, while the slow flow possibly arises from the outer Lagrange point as the nova envelope first encased the binary orbit.}
\label{fig:illustration}
\end{figure*}

High-spatial resolution observations performed using the Very Large Telescope with VISIR from 2009 to 2011 \citep{2012AA...545A..63C} and using the Gemini South telescope with T-ReCS from 2010 to 2012 \citep[][also see Figure \ref{fig:V1280Sco_Gemini_final}]{2016ApJ...817..145S} revealed a bipolar-shaped (or elongated shape) dusty nebula round V1280 Sco (see Figure \ref{fig:LC_SMEI_V_final} for the observation dates).
\citet{2012AA...545A..63C} suggested that the mass-loss was dominantly polar.
Besides, \citet{2013PASJ...65...37N} described the structure of the ejected shell as in the left of Figure \ref{fig:illustration}, where the inclination angle was supposed to be low.
Their picture was based on the presence of the dusty bipolar nebula and multiple absorbing gases (Na\,{\footnotesize I} D, Ca\,{\footnotesize II} H and K, and metastable He\,{\footnotesize I}* ), velocity range of which is from $-650$ \kms\, to $-900$ \kms, moving on the line of sight detected by the Subaru telescope with HDS.
Whereas, as shown in Section \ref{sec:modeling_results}, [O\,{\footnotesize III}]  and  [N\,{\footnotesize II}] forbidden line regions ($V_{\rm exp} = 1600-2100$ \kms, fast outflow) are expanded/elongated with an inclination angle of $\sim$80$^{\circ}$, that is, almost vertically with the line of sight.
These findings enable us to improve the drawing of V1280 Sco as in the right of Figure \ref{fig:illustration}, where the inclination angle is high (edge-on) and the ejecta consist of an equatorial torus (ring) and a bipolar outflow.
Within the slower component of the ejecta, the density is often enhanced toward the equatorial plane.
This equatorial torus may shape the faster component of the ejecta into a bipolar morphology \citep[e.g.,][]{2016acps.confE..21S}.
This idea is consistent with the radio or optical direct imaging studies of some novae such as V959 Mon \citep[$\gamma$-ray nova,][]{2015ApJ...805..136L, 2017MNRAS.469.3976H}, T Pyx \citep[recurrent nova,][]{2011AA...534L..11C, 2016acps.confE..21S}, and HR Del \citep[slow nova,][]{2009AJ....138.1541M, 2003MNRAS.344.1219H}, which reveal that the ejecta from novae appear to consist of two main components: a slow, dense outflow and a fast outflow or wind.

\subsection{Eruption scenario}
\label{sec:ejection scenario}
Since the first detection of GeV $\gamma$-ray emission in a symbiotic nova V407 Cyg with the Large Area Telescope aboard the {\it Fermi Gamma-Ray Space Telescope} in 2010 \citep{2010Sci...329..817A}, $\gamma$-rays have been detected  ($>$3$\sigma$ significance) from over a dozen Galactic novae and an existence of a correlation between the optical and $\gamma$-ray light curves, at least in some novae (but clearly in V959 Mon and V5856 Sgr), is argued \citep[e.g.,][]{2017NatAs...1..697L, 2020NatAs...4..776A}.
In order to explain observational results of novae represented by the detection of $\gamma$-rays, an internal radiative shock model (hereafter, shock model) has been proposed, where the shocks take place in the orbital plane of the binary system when an initial slow ($\sim$ a few hundred \kms), equatorially focused outflow is collided from behind by a fast flow ($\sim$ a few thousand \kms), more spherically symmetric outflow \citep[e.g.,][]{2014Natur.514..339C, 2020arXiv201108751C, 2015MNRAS.450.2739M, 2017NatAs...1..697L, 2020NatAs...4..776A, 2020ApJ...905...62A}.
While there are no reports indicating the detection of $\gamma$-ray at the early phase of V1280 Sco, it shows the common features expected from the shock model; particularly, optical flares near the maximum light and dust formation \citep[][also see Figure \ref{fig:LC_SMEI_V_final}]{2012AA...543A..86N}.
The morphology is crucial to discuss if the shock model is applicable to the extremely slow nova V1280 Sco.
The extremely slow and peculiar nova V1280 Sco and the $\gamma$-ray bright nova V959 Mon are likely to share common characteristics of slow, equatorial torus (ring) and faster, bipolar outflows.
\citet{2014Natur.514..339C} used radio observations of V959 Mon to suggest that collisions between a dense equatorial torus and a faster flow led to shocks that accelerated particles, explaining that many normal novae produce GeV $\gamma$-ray.
Here we discuss an eruption scenario for V1280 Sco based on the model by \citet{2020ApJ...905...62A}, which are proposed by analysing a sample set of 12 novae including $\gamma$-ray non-detected ones.
In the beginning of eruption, the accreted envelope puffs up due to the energy of the thermonuclear runaway, and engulfs the binary system producing a common envelope.
The envelope is likely to become concentrated in the orbital plane through the outer Lagrange point \citep{2016MNRAS.455.4351P}, which may be strongly affected by the orbital motion \citep[e.g.,][]{1990ApJ...356..250L} and/or some other mechanisms like magnetic field \citep[e.g.,][]{2011AA...536A..97F}.
At this stage, P Cygni like absorptions with velocity of a few hundred \kms, associated with a slow flow, are observed. 
In the case of V1280 Sco, a typical velocity of the slow flow was measured to be 350 $\pm$ 160 \kms\, using O\,{\footnotesize I} and Si\,{\footnotesize II} lines in low-dispersion spectra near the optical peak \citep{2012AA...543A..86N}.
The slow flow is followed by a faster wind, which propagates more freely in the bipolar directions due to the pre-existing slower ejecta concentrated in the equatorial plane.
The continuous fast wind, however, driven by radiation from the ongoing nuclear burning on the surface of the WD may be spherical rather than bipolar \citep{1994ApJ...437..802K}.
As the fast wind catches up with the slow flow, a shock interaction is expected to occur and accelerate the slow flow to an intermediate velocity as referred in \citet{2020ApJ...905...62A}.
The dust can form in the compressed (dense) and radiatively cooled region, which is produced by the collision between the slow flow and the fast wind \citep{2017MNRAS.469.1314D}.
On V1280 Sco, it is reasonable to suppose that the velocities of a slow, intermediate, and fast flow correspond to  $\sim$350 \kms, $650-900$ \kms, and $\sim$2000 \kms, respectively.
The observational fact that V1280 Sco showed three short episodes of brightening (each time scale is less than one day) with amplitudes of $\sim$1 mag near the peak \citep{2010ApJ...724..480H}, a deep extinction due to dust formation $\sim$10 days after the maximum \citep{2012AA...543A..86N}, and multiple intermediate-velocity clumpy components moving to the line of sight until several years after the peak \citep{2010PASJ...62L...5S, 2013PASJ...65...37N}, is consistent with a high-inclination, that is, an edge-on binary system if the shock model is adopted.
The unique light curve as shown in V1280 Sco is also probably to be related to its high-inclination.

\subsection{Distance and peak luminosity}
\label{sec:distance and peak luminosity}
The distance to V1280 Sco was estimated in various approaches.
\citet{2010ApJ...724..480H} estimated the distance to be $d$ = 630 $\pm$ 100 pc, where they obtained the peak luminosity ($L_{\star}$) as
\begin{equation}
L_{\star} = 2.4 \times 10^{4} (\frac{t_{\mathrm{c}}}{24\,\mathrm{days}})^{2}(\frac{v_{\mathrm{ej}}}{600\, \mathrm{km}\,\mathrm{s}^{-1}})^{2}(\frac{T_{\mathrm{c}}}{1200\, \mathrm{K}})^4 L_{\sun},
	\label{eq:eq2}
\end{equation}
by measuring the condensation time ($t_{\mathrm{c}}$ = 24 days) of dust grains from SMEI data (see Figure \ref{fig:LC_SMEI_V_final}), assuming that the condensation temperature ($T_{\mathrm{c}}$) was 1200 K and the ejection velocity ($v_{\mathrm{ej}}$) was 600 \kms, and then compared it with the maximum light $m_{V}^{\mathrm{max}} = 4$, taking the interstellar extinction $A_{V} = 1.2 \pm 0.3$.
On the other hand, \citet{2008AA...487..223C} and \citet{2012AA...543A..86N} derived $d$ = 1.6 $\pm$ 0.4 kpc and $d$ = 1.1 $\pm$ 0.5 kpc from the expansion parallax of a dusty shell assuming the velocity is 500 \kms\, and 350 \kms, respectively, where the difference depends on taking account of the flowing direction and velocity field of the dusty ejecta; for example the morphology is changeable from a torus-dominant to a bipolar-shaped geometry as suggested for V339 Del \citep{2019ApJ...872..120K}.
Furthermore, \citet{2016ApJS..223...21H} obtained the distance as 0.96 kpc by examining the $UBV$ color evolution based on their universal decline law model, which is consistent with our estimation of  1.1 kpc.
Here we attempt to estimate the distance by using Equation (\ref{eq:eq2}) in a similar way to \citet{2010ApJ...724..480H}.
When we adopt 350 $\pm$ 160 \kms\, instead of 600 \kms\, as the ejection velocity ($v_{\mathrm{ej}}$), and assume that the magnitude at the premaximum halt ($\sim$6.3 mag; $\sim$2.5 mag larger than the maximum) corresponds to its Eddington luminosity for $\sim$0.6 $M_\mathrm{\sun}$ WD \citep{2004ApJ...612L..57H}, a distance of 1.06 $\pm$ 0.49 kpc is derived.
Hence, we may conclude that the mass of WD in V1280 Sco system is low ($\sim$0.6 $M_\mathrm{\sun}$) and  its distance is $\sim$1.1 kpc, and the peak magnitude is  $\sim$2.5 mag super-Eddington.

The MMRD relationship is frequently used to estimate distances to Galactic novae especially for statistical studies, while this is not completely explained in theory.
Recently, \citet{2020ApJ...902...91H} clarify the physics of MMRD points by considering the mass accretion rate onto the WD along with the optically thick wind theory, and conclude that the MMRD relation is not to be a precision distance indicator of individual nova due to its irrepressible scatter. 
The scatter may be contributed from the novae with their luminosities powered by shock interactions correspondingly.
Slow novae tend to occur interaction between ejecta like shown in V1280 Sco, which can cause the deviation from the MMRD relation.
For measurement of the distance to a nova, \citet{2018MNRAS.481.3033S} recommend,  in order of accuracy, (1) using the {\it Gaia} parallax and (2) using the catalogue of \citet{2016MNRAS.461.1177O} where distances are calculated using their specific reddening-distance relations.
\citet{2016MNRAS.461.1177O} provided the calculated distance for V1280 Sco as 1.4 $\pm$ 0.9 kpc, which is in good agreement with our estimation within uncertainties.
According to {\it Gaia} Early Data Release 3\footnote{Available from https://gea.esac.esa.int/archive/}, however, the parallax corresponding to $3.64^{+0.27}_{-0.24}$ kpc is released, whereas it is substantially larger than any previous estimation.
If the {\it Gaia}'s value is adopted, the maximum absolute magnitude of V1280 Sco is estimated to be $M_{V} = -10.23^{+0.13}_{-0.14}$, which is extremely brighter than the typical slow nova \citep[e.g.,][]{2019AA...622A.186S}.
Future more observations are needed to determine the distance to V1280 Sco and to discuss the difference of the estimated distance among them.

\begin{deluxetable*}{llccccccc}
\tablecaption{Examples of novae with morpho-kinematic modeling with \textsc{\lowercase{SHAPE}}.}
\label{table:novae_list}
\tablehead{
\colhead{Name (outburst year)} & \colhead{Morphology} & \colhead{$i$ (deg)} & \colhead{$\Delta{t}$}\tablenotemark{a}  & \colhead{Analized line(s)} & \colhead{Imaging}\tablenotemark{b} & \colhead{Ref.}
}\
\startdata
V1280 Sco (2007)  &  Bipolar structure (and equatorial torus/ring) & $81^{+2}_{-4}$ & $+1589$ d &  [N\,{\footnotesize II}] 5755 & (Gemini-S) & 1 \\
 & Bipolar structure (and equatorial torus/ring) & $80^{+1}_{-3}$ & $+1846$ d &  [O\,{\footnotesize III}] 4959, 5007 &  (Gemini-S)  & 1 \\
T Aur (1891) & A peanut shaped shell & $75 \pm 2$ & $+125.0$ y & H${\alpha}$ (position-velocity) & NOT\tablenotemark{c}  & 2\\
DQ Her (1934) & A prolate ellipsoid  & $87 \pm 2$ & $+82.4$ y & H${\alpha}$ (position-velocity) & NOT\tablenotemark{c}  & 2\\
HR Del (1967) & A prolate ellipsoid & $37 \pm 3$ & $+53.1$ y & H${\alpha}$ (position-velocity) & NOT\tablenotemark{c}  & 2\\
& with an equatorial component & & &  & & \\
QU Vul (1984) & A spherical shape &  & $+35.6$ y & H${\alpha}$  (position-velocity) & NOT\tablenotemark{c}  & 2\\
RS Oph (2006) & Dumbbell structure& $39^{+1}_{-10}$ & $+155$ d&  [O\,{\footnotesize III}] 5007 & {\it HST} & 3\\
 & with an hour-glass overdensity & &  &  & &  & \\
V2491 Cyg (2008) & Polar blobs and an equatorial ring & $83^{+3}_{-12}$ & $+108$ d &  H${\alpha}$ + [N\,{\footnotesize II}] 6548, 6584 & & 4\\
V2672 Oph (2009) & Prolate system & $0 \pm 6$ & $+8.33$ d & H${\alpha}$ &  & 5\\
 & with polar blobs and an equatorial ring& & &   & & \\
KT Eri (2009) & Dumbbell structure & $58^{+6}_{-7}$ & $+42.29$ d & H${\alpha}$ & & 6\\
V959 Mon (2012) & Bipolar structure (and equatorial ring) & $82 \pm 6$ & $+130$ d &  [O\,{\footnotesize III}] 4959, 5007 & ({\it HST}) & 7, 8\\
V5668 Sgr (2015) & An equatorial disk & 85 & $+822$ d &  [O\,{\footnotesize III}] 4363 & ({\it HST}, Keck) &  9, 10\\
 & An equatorial waist and polar cones & 85 & $+822$ d &  [O\,{\footnotesize III}] 4959, 5007 &  ({\it HST}, Keck)  &  9, 10\\
V906 Car (2018) & Asymmetric bipolar structure & $53 \pm 1.55$ & $+21$ d & H${\alpha}$ & & 11\\
 & Asymmetric bipolar structure & $60 \pm 1.75$ & $+21$ d & O\,{\footnotesize I} 8446 & & 11\\
 & Asymmetric bipolar ellipsoidal-like structure& $50 \pm 1.25$  & $+96$ d & [N\,{\footnotesize II}] 5755 & & 11\\
 & Asymmetric bipolar structure& $50 \pm 1.95$ & $+316$ d & H${\alpha}$ & & 11\\
 &  along  with equatorial rings & & & & &\\
 & Asymmetric bipolar and triangular polar ends & $35 \pm 1.65$ & $+316$ d & He\,{\footnotesize I} 5876 & & 11\\
\enddata
\tablenotetext{a}{Days or years after outburst defined in each reference for which data was used to analyze the inclination angle.}
\tablenotetext{b}{Telescopes used for high-resolved imaging observations. The parentheses indicate that the references for the morpho-kinematic modeling and the high-resolved imaging are different.}
\tablenotetext{c}{The 2.56 m Nordic Optical Telescope (NOT) at the Roque de los Muchachos Observatory (ORM, La Palma, Spain).}
\tablerefs{(1) \cite{2016ApJ...817..145S}; (2) \citet{2022MNRAS.512.2003S}; (3) \citet{2009ApJ...703.1955R}; (4) \citet{2011MNRAS.412.1701R}; (5) \citet{2011MNRAS.410..525M}; (6) \citet{2013MNRAS.433.1991R}; (7) \citet{2013ApJ...768...49R}; (8) \citet{2016acps.confE..21S}; (9) \citet{2018AA...611A...3H}; (10) \citet{2022MNRAS.511.1591T}; (11) \citet{ 2020MNRAS.495.2075P}}
\end{deluxetable*}

\subsection{V1280 Sco and other novae with morpho-kinematic modeling}
\label{sec:comparison between V1280 Sco and other novae with morpho-kinematic modeling}
Morpho-kinematic modeling provides information on the expansion velocity, inclination angle and the morphology of the ejected shell following a nova outburst.
The technique has been applied to novae whose resolved images and/or spectroscopic line profiles are observed.
Simultaneous resolved imaging and spectroscopic observations complementary disentangle the morphology and kinematics of the ejecta.
While resolved images have the advantage that they reveal a myriad of structures, they are not obtained until the ejecta expand enough to be resolved.
This means that the information on morphology of the ejecta at the early outburst phase tends not to be available.
On the other hands, spectroscopic observations allow us to constrain the model, including the change of the morphology, from the early phase to later stages.
However, it is difficult to model the ejecta without the aid of imaging for the case that spectral line profiles are composed of multiple and complex components.

Table \ref{table:novae_list} shows examples of novae with morpho-kinematic modeling with \textsc{\lowercase{SHAPE}}.
In this paper we modeled V1280 Sco over $\sim$1500 days after outburst as a bipolar structure of forbidden lines of [O\,{\footnotesize III}] 4959, 5007 and [N\,{\footnotesize II}] 5755, along with an equatorial torus (ring) based on the high resolution image of dusty nebula with the Gemini South/T-ReCS and the detection of multiple absorption lines on the line of sight.
Another most successful work, where the interplay between resolved imaging and ground based spectroscopic observations play a great role, was performed for RS Oph \citep{2009ApJ...703.1955R}.
{\it Hubble Space Telescope} ({\it HST}) ACS/HRC narrow band (F502N filter) imaging and ground based spectroscopic observations ([O\,{\footnotesize III}] 5007 line profile) of RS Oph at 155 days after outburst drew a shape of ejecta as a bipolar composed of an outer dumbbell and inner hour glass structures.
However, the shape can be constructed in only one phase even though such a high quality data set is used.
Actually, according to \citet{2009ApJ...703.1955R}, the [O\,{\footnotesize III}] 5007 line of RS Oph has changed from narrow to wide, suggesting that narrow emission lines do not originate from the expanding remnant alone but arises from the ionized red giant (companion star) wind ahead of the forward shock.
It is harder to draw firm conclusions unless there are simultaneous imaging and spectroscopy.
In that respect, V1280 Sco has a potential to be observed with simultaneous spatially resolved imaging and high-dispersion spectroscopy in the near future due to its extremely slow evolution, which may make it possible to give new findings on evolution of morphological and spectral profiles as well as their correlation.
This expectation is based on that V1280 Sco shows the following similar results with novae V959 Mon and V5668 Sgr \citep{2013ApJ...768...49R, 2018AA...611A...3H}.

Despite of the different property between V1280 Sco (dust nova, long plateau) and V959 Mon ($\gamma$-ray nova, smooth decline), they are modeled alike as a bipolar structure with $i$ = $\sim$80 deg using [O\,{\footnotesize III}] 4959, 5007 profiles on 1846 days and on 130 days after outburst, respectively.
Note that the images of V959 Mon were obtained with {\it HST} and the WFC3 camera (F657N and F502N filters) in 2014 and 2015, which are corresponding to approximately 2.5 and 3.5 years after outburst, where F657N filter traces mainly dense gas of H$\alpha$ while F502N filter traces more diffuse gas of [O\,{\footnotesize III}] \citep{2016acps.confE..21S}.
The 2014 {\it HST} images indicate that [O\,{\footnotesize III}] emission appeared in bipolar shape which is roughly consistent with the morphology inferred by \citet{2013ApJ...768...49R}, whereas dense, edge-on equatorial torus was dominantly shown only in F657N (H$\alpha$ +  [N\,{\footnotesize II}]) filter.
On the other hand, the 2015 {\it HST} images indicate that H$\alpha$ and [O\,{\footnotesize III}] emissions from the outer, fast flow had almost completely faded and the inner spherical  [O\,{\footnotesize III}] shell took on the appearance of two arcs.
These changes can be reflected in the spectral feature naturally, leading to that such spectral profile may be modeled as a bipolar structure and an equatorial ring as shown in V5668 Sgr.
If it is the case of V1280 Sco, the [O\,{\footnotesize III}] profile may change to be contributed from the emission emanated from the equatorial torus as the system evolves (the density distribution and/or the ionization state change).
Furthermore, the high inclination angle (edge-on) of V1280 Sco may be used as a predictor for searches of eclipses which will then provide us further information on the system parameters, such as the orbital period and the white dwarf mass.
In fact, a orbital period of 7.1 hr was observed partial eclipse from extended emission by an accretion desk rim in V959 Mon \citep{2013ApJ...768L..26P}.

\subsection{Observability of the [O\,{\footnotesize III}] direct imaging}
\label{sec:Observability of  OIII and NII images}
In this section we discuss the observability of the [O\,{\footnotesize III}] direct imaging with the {\it HST} and the Wide Field Channel 3 (WFC3), with a scale of 0".04 pixel$^{-1}$ as an example.
V1280 Sco keeps the brightness of $\sim$10 mag ($V$) which is contributed from the free-free continuum radiation and the emission lines at comparable levels.
According to the latest high-dispersion spectrum obtained on 2019 Jun 12, the ratio of the total continuum flux to the [O\,{\footnotesize III}] 5007 emission flux in the range of F502N narrowband filter (full width at 10\% of peak transmission: 57.8 \AA) is $\sim$3. 
If we adopt an expansion velocity of $\sim$2100 km/s, an inclination of $\sim$80$^{\circ}$, and a distance of $\sim$1.1 kpc, the [O\,{\footnotesize III}] region should expand to $\sim$10 arcseconds in apparent major axis size in 2022 (15 years after outburst).
Therefore, we can estimate the exposure time for the {\it HST} observation of  [O\,{\footnotesize III}] region assuming $\sim$16.2 mag per square arcseconds (corresponding to $\sim$22.7 mag per pixel size).
When using the {\it HST} and the WFC3 with the F502N filter, an image of  [O\,{\footnotesize III}] region with a signal-to-noise  (S/N) ratio per pixel of 10 can be obtained with $\sim$300 seconds exposure time.
However, the difficulty for the  [O\,{\footnotesize III}] direct imaging (without a coronagraphic mask) on V1280 Sco is due to the central bright point source ($V \sim$10). A total of about 60 frames with each 5.2 seconds exposure time is needed for S/N = 10 without saturating the central source.
We also considered the requirements for follow observations with space based telescopes.

\section{Conclusions}
\label{sec:conclusions}

In this paper we studied the morphology of the expanding ejecta of the extremely slow nova V1280 Sco with a unique light curve.
We compare synthetic line profile spectra to the observed spectra for [O\,{\footnotesize III}] 4959, 5007 and [N\,{\footnotesize II}] 5755 emission lines in order to find the best-fit morphology, inclination angle, and maximum expansion velocity of the ejected shell.
We derive the best fitting expansion velocity, inclination, and squeeze as $V_{\rm exp} = 2100^{+100}_{-100}$ \kms, $i = 80^{+1}_{-3}$ deg, and $squ = 1.0^{+0.0}_{-0.1}$ using [O\,{\footnotesize III}] line profiles, and $V_{\rm exp} = 1600^{+100}_{-100}$ \kms, $i = 81^{+2}_{-4}$ deg, and $squ = 1.0^{+0.0}_{-0.1}$ using  [N\,{\footnotesize II}] 5755 line profile, respectively.
A high inclination angle is consistent with the observational results of showing multiple absorption lines originating from clumpy materials which are likely to be produced in a dense and slow equatorially focused outflow.
The observational fact that V1280 Sco showed three short episodes of brightening with amplitudes of $\sim$1 mag near the peak, a deep extinction due to dust formation $\sim$10 days after the maximum, and multiple intermediate-velocity clumpy components moving to the line of sight until several years after the peak, is consistent with an edge-on binary system if the internal shock model is adopted.
We discuss the distance to V1280 Sco and suggest that the mass of WD is low ($\sim$0.6 $M_\mathrm{\sun}$) and  its distance is $\sim$1.1 kpc, and the peak magnitude is  $\sim$2.5 mag super-Eddington.
Considering V1280 Sco is the extremely slow nova, future more observations are able to be conducted to investigate the nature of this nova.
Growing the sample size of novae for which morphology information is available will be helpful for addressing long-standing mysteries for nova outbursts such as the dominant energy source to power the optical light at the maximum, optical flares near the maximum, and dust formation.



\newpage

\begin{acknowledgments}
This work is based on data collected at the Subaru Telescope, which is operated by the National Astronomical Observatory of Japan.
The Pirka telescope is operated by Graduate School of Science, Hokkaido University, and is partially supported by the Optical and Near-Infrared Astronomy Inter-University Cooperation Program, MEXT, of Japan.
This work (partially) has made use of data from the European Space Agency (ESA) mission {\it Gaia} (\url{https://www.cosmos.esa.int/gaia}), processed by the {\it Gaia} Data Processing and Analysis Consortium (DPAC, \url{https://www.cosmos.esa.int/web/gaia/dpac/consortium}).
Funding for the DPAC has been provided by national institutions, in particular the institutions participating in the {\it Gaia} Multilateral Agreement.
We are grateful to students at Osaka Kyoiku University for providing the photometric data that we used in this research.
We thank Nayoro Observatory staff for their supports.
Thanks are also due to Izumi Hachisu and Mariko Kato for their useful comments.
The publication of this work is financially supported by National Astronomical Observatory of Japan.
V.A.R.M.R. acknowledges financial support from the South African Square Kilometre Array Project for the postdoctoral fellowship position at the University of Cape Town, the Radboud Excellence Initiative, the Funda\c{c}\~ao para a Ci\^encia e a Tecnologia (FCT) in the form of an exploratory project of reference IF/00498/2015/CP1302/CT0001, FCT and the Minist\'erio da Ci\^encia, Tecnologia e Ensino Superior (MCTES) through national funds and when applicable co-funded EU funds under the project UIDB/EEA/50008/2020, and supported by Enabling Green E-science for the Square Kilometre Array Research Infrastructure (ENGAGE-SKA), POCI-01-0145-FEDER-022217, and PHOBOS, POCI-01-0145-FEDER-029932, funded by Programa Operacional Competitividade e Internacionaliza\c{c}\~a o (COMPETE 2020) and FCT, Portugal.
H.N. and V.A.R.M.R. acknowledge financial support from the Global COE Program of Nagoya University ”Quest for Fundamental Principles in the Universe (QFPU)” from JSPS and MEXT of Japan in the form of a visiting scholar fellowship.
\end{acknowledgments}

%

\vspace{5mm}
\facilities{Subaru (HDS), Gemini:South (T-ReCS), Pirka:1.6m (MSI), Kamogata/Kiso/Kyoto Wide-field Survey, {\it Gaia}}


\software{ 
          IRAF \citep{1986SPIE..627..733T}, SHAPE \citep{2011ITVCG..17..454S}
          }





\bibliography{V1280Sco_III}{}
\bibliographystyle{aasjournal}


\listofchanges

\end{document}